\documentclass[
aps,
prl,
twocolumn
]{revtex4-2}

\usepackage{amsmath}
\usepackage{amssymb}
\usepackage{graphicx} 
\usepackage{dcolumn}  
\usepackage{bm}       
\usepackage{color}     
\usepackage[mathlines]{lineno}

\begin{document}
	
	\preprint{APS/123-QED}
	
	\title{Thermalization and irreversibility of an isolated quantum system}
	\author{Xue-Yi Guo}
	\email{guoxy@baqis.ac.cn}
	\affiliation{
		Beijing Academy of Quantum Information Sciences, Beijing 100193, China
	}
	
	\date{\today}

\begin{abstract}
The irreversibility and thermalization of many-body systems can be attributed to the erasure of spread non-equilibrium state information by local operations. This thermalization mechanism can be demonstrated by the sequence of $\hat{O}_i(t_i)$, where $\hat{O_i}$ is a local operator, $\hat{O_i}(t_i) = e^{i\hat{H}t_i} \hat{O_i} e^{-i\hat{H}t_i}$, $\hat{H}$ is the system Hamiltonian, $t_i$ can take positive, negative, or zero values, and the sequence is arranged according to the subscript $i$.
We numerically demonstrate the information erasure of initial non-equilibrium quantum states through such sequence in a one-dimensional Hubbard model. During this process, the system's entanglement entropy increases monotonically toward a stable value.
By incorporating this information erasure mechanism into an isolated system, our numerical simulations reveal that in this completely isolated system, a thermalization process emerges.

\end{abstract}

\maketitle

\textit{Introduction}.— 
According to the time-reversal symmetry of the Schrödinger equation, all microscopic dynamical processes are reversible. However, numerous natural phenomena exhibit irreversibility, with thermodynamic processes being a prime example—for instance, when a hot gas and a cold gas mix, they eventually reach thermal equilibrium at an intermediate temperature. This seeming paradox has led to extensive theoretical investigations.
Von Neumann explained that the principles of statistical mechanics could be understood within the framework of quantum mechanics by invoking the ergodicity of quantum dynamical processes \citep{von_neumann_proof_2010}. Deutsch and Srednicki proposed the eigenstate thermalization hypothesis (ETH)\citep{deutsch_quantum_1991, srednicki_chaos_1994}, which states that the high-energy eigenstates of quantum systems with complex internal interactions, such as gases, intrinsically exhibit thermal properties. Thermalization can also be studied from the perspectives of quantum information scrambling and quantum chaos \citep{hayden_black_2007, shenker_black_2014, Kitaev_alexei_2015, maldacena_bound_2016}.

Theoretically, in an isolated system, regardless of system size or interaction complexity, if the Hamiltonian $\hat{H}$ can be inverted as $-\hat{H}$, the system evolves backward to its initial state, preserving quantum information.  
Now, let us consider a thought experiment: Initially, gas particles are confined to a container's corner, with each wave function forming a localized wave packet, exhibiting low position uncertainty and high momentum uncertainty.  
As the wave packets evolve over time and spread in space, collisions and scattering among the particles lead to a near-uniform state, where each particle becomes highly entangled with other particles and exhibits significant position uncertainty.  
To reverse the evolution and bring the gas particles back to their initial localized state, the system would have to disentangle the highly entangled wave functions through a series of collisions and reconstruct the original localized wave packets via constructive interference. In fact, from a quantum mechanical perspective, any process that leads to a decrease in thermodynamic entropy generally involves disentanglement and constructive interference. 
Experimentally, achieving such a reversal imposes extremely stringent conditions. Even small local perturbations $\hat{O}$ can disrupt the disentanglement and constructive interference process. In our thought experiment, interactions between the gas particles and the container walls could introduce slight perturbations, altering the phase along specific transmission paths and disrupting the coherence required for perfect reversal. 

One approach to estimating the impact of local perturbations on the reversibility of a quantum system is to observe the system's Loschmidt Echo\citep{peres_stability_1984, schmitt_effective_2016, schmitt_irreversible_2018, schmitt_semiclassical_2019},
\begin{equation}
	L(t) = \left| \langle \psi_0 | e^{i H t} e^{-i (H + \delta H) t} | \psi_0 \rangle \right|^2.
\end{equation}
$|\psi_0\rangle$ is the initial quantum state, $ H $ is the unperturbed Hamiltonian, and $\delta H $ represents a small perturbation to the system. The Loschmidt Echo measures the overlap between the time-evolved states under $ H $ and $ H + \delta H $, capturing the reversibility of quantum dynamics and the stability of quantum information. A rapid decay of $ L(t) $ indicates strong sensitivity to perturbations, which is often associated with quantum chaos and thermalization.
Here, we focus on the information erasure effect of the sequence of $\hat{O_i}(t_i)$. 
For a typical macroscopic system, the operation $\hat{O_i}(t_i)$ is expected to erase the spread information of the system's initial non-equilibrium state, thereby preventing it from returning to non-equilibrium.
Consequently, the application of the sequence of $\hat{O_i}(t_i)$ should lead to a monotonic increase in the system's entanglement entropy, representing a mechanism for irreversibility and thermalization.
The introduction of such perturbations may seem to violate the assumption of an isolated system. However, this issue can be resolved by incorporating the perturbation source, such as the container walls in the thought experiment, into the isolated system. 
Here, $\hat{O_i}(t_i) = e^{i\hat{H}t_i} \hat{O_i} e^{-i\hat{H}t_i}$ can be regarded as the time evolution or propagation of the operator $\hat{O_i}$ under the Hamiltonian $\hat{H}$. A similar sequence is the out-of-time-order correlator (OTOC), which serves as an important tool for studying the dynamical evolution of quantum information and the thermalization properties of quantum many-body systems \citep{swingle_unscrambling_2018, murthy_bounds_2019, chan_eigenstate_2019, foini_eigenstate_2019, brenes_out--time-order_2021}. \textcolor{red}{\textit{The essence of the OTOC sequence lies in the emergence of non-commutativity between initially commuting local operators after time evolution.}}

A simplified gas-like quantum many-body system can be constructed using the Hubbard model. The Hubbard model discretizes space using a lattice, with particles propagating through space via hopping between lattice sites.
Potential energy variations in space can be implemented through site-dependent potential terms, while interactions between particles—such as collision and scattering—are modeled using interaction terms between particles. The Hubbard model has been used to study thermalization phenomena, including investigations of the dynamical properties of the system under different Hubbard model parameters \citep{kollath_quench_2007}, the thermalization dynamics following a quench \citep{eckstein_thermalization_2009, sorg_relaxation_2014}, and the thermalization of subsystems within the Hubbard model \citep{genway_dynamics_2010}.
With recent advancements in quantum simulation and quantum computing technologies, experimental studies of thermalization phenomena have been conducted in both cold-atom systems and superconducting quantum processors. In optical lattices, cold-atom systems based on Bose-Einstein condensates have been used to explore entanglement and thermalization \citep{kaufman_quantum_2016}. Meanwhile, superconducting quantum chips have facilitated studies on ergodic dynamics, thermalization, and information scrambling in isolated quantum systems \citep{neill_ergodic_2016, chen_observation_2021, zhu_observation_2022}.

In this work, we construct a quantum many-body system in which the initial state information rapidly spreads. Numerical simulations demonstrate the fast-growing entanglement entropy dynamics starting from a non-equilibrium state. We then numerically show the information erasure and the resulting irreversible thermalization effects induced by the sequence of $\hat{O_i}(t_i)$ in this system. By introducing the corresponding local perturbations into the model, we construct an isolated quantum many-body system that exhibits rapid thermalization. Finally, we discuss the feasibility of conducting related research on superconducting quantum computing chips.

\textit{Model}.— 
To model gas particle transport and collisions, we define a quantum many-body system on a lattice, which can be one-, two-, or three-dimensional. Particles, assumed to be fermions, hop between adjacent sites, with at most one particle of each type per site. Different particle types can coexist and interact attractively or repulsively. The system is governed by the Hamiltonian:

\begin{align}\label{hamiltonian}
	H &= \sum_{\tau}\sum_{\langle i,j \rangle} J_\tau \left(c_{i,\tau}^\dagger c_{j,\tau} + c_{i,\tau} c_{j,\tau}^\dagger \right) \notag \\
	&\quad + \sum_{\tau,i} U_{i, \tau} n_{i,\tau} \notag \\
	&\quad + \sum_{i} \sum_{\underset{\tau \neq \upsilon}{\tau,\upsilon}} U_{\tau,\upsilon} n_{i,\tau} n_{i,\upsilon}.
\end{align}

\begin{figure}[htbp]  
	\centering  
	\includegraphics[scale=1.0]{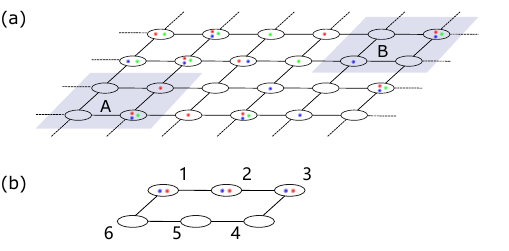}   
	\caption{Schematic representation of the simplified gas-like model. The circles denote lattice sites, while the connecting lines indicate possible hopping between neighboring sites. The small spheres represent particles, with different colors distinguishing different particle types.
(a) A two-dimensional lattice system containing three types of fermions: red, blue, and green.
(b) A one-dimensional ring lattice with a length of six sites. The figure shows an initial non-equilibrium Fock state, where red ($\tau$) and blue ($\upsilon$) particles are placed at sites 1–3. }  \label{fig1}  
	
\end{figure}

Here, $ i, j $ are site indices, $\langle i,j \rangle$ represents nearest-neighbor lattice sites, and $ \tau, \upsilon $ label particle types. The operators $ c_{i,\tau}^\dagger $ and $ c_{i,\tau} $ represent the creation and annihilation of a particle of type $ \tau $ at site $ i $, respectively. The parameter $ J_\tau $ denotes the hopping amplitude for particles of type $ \tau $ between neighboring sites, while $ U_{i,\tau} $ represents the local potential energy of particle $ \tau $ at a given site $i$. The interaction strength between particles of different types at the same site is given by $ U_{\tau,\upsilon} $.  
In this model, the interaction strength $ U_{\tau,\upsilon} $ is comparable to the hopping amplitude $ J_\tau $, meaning that different particle types can move freely across the lattice without becoming strongly bound together. This model can also be extended to describe bosonic particles.  
A similar parameter setup has been employed in previous studies \citep{genway_dynamics_2010, genway_thermalization_2012}, and \citep{kollath_quench_2007} has investigated how different parameter regimes affect the non-equilibrium dynamical evolution of the system.

By structuring the lattice and tuning interactions, we control particle transport and collisions while exploring entanglement between particle types and spatial regions. Fig.~\ref{fig1}(a) depicts a two-dimensional lattice hosting three fermion types, where circles represent sites, and colored spheres denote different fermions. Lines indicate hopping between neighboring sites. 
In a system like Fig.~\ref{fig1}(a), thermal equilibrium is defined by the following criterion: for any two subsystems, A and B, much smaller than the total system, local observables must have identical expectation values and fluctuations. Moreover, A and B should exhibit no classical or quantum correlations, implying zero mutual information.

To reduce the complexity of numerical simulations, we consider a one-dimensional model with six lattice sites, as shown in Fig.~\ref{fig1}(b). The one-dimensional chain is subject to periodic boundary conditions, and it contains two types of fermions that can propagate and interact within the lattice. The Hamiltonian of this model is given by:
\begin{align}
    H &= \sum_{i = 1}^6 J_\tau \left(c_{i,\tau}^\dagger c_{i+1,\tau} + c_{i,\tau} c_{i+1,\tau}^\dagger \right) \notag \\
    &\quad + J_\upsilon \left(c_{i,\upsilon}^\dagger c_{i+1,\upsilon} + c_{i,\upsilon} c_{i+1,\upsilon}^\dagger \right) \notag \\
    &\quad + U_{\tau,\upsilon} n_{i,\tau} n_{i,\upsilon} .
\end{align}
For $ i=6 $, we impose periodic boundary conditions such that $ i+1 \equiv 1 $. The hopping amplitudes for both particle types are set to the same value, $ J_{\tau(\upsilon)} = 1 $, and the on-site interaction potential between different particle types is given by $ U_{\tau,\upsilon} = -0.05 $.

\begin{figure}[htbp]  
	\centering  
	\includegraphics[scale=1.0]{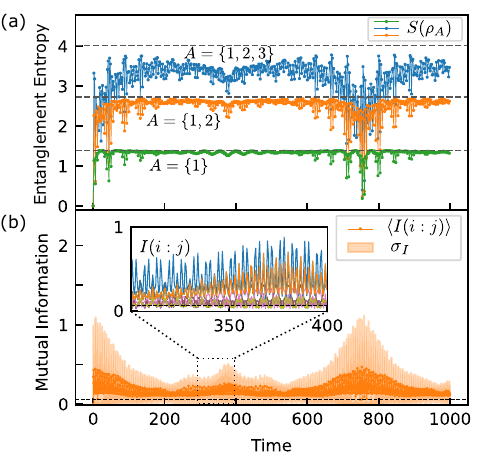}  
	\caption{The dynamical evolution of entanglement entropy and mutual information.  
		(a) The evolution of entanglement entropy $S_{\rho}$ is presented for different subsystem sizes $A$. The blue, orange, and green lines correspond to $A = \{1,2,3\}$, $\{1,2\}$, and $\{1\}$, respectively. The black dashed line denotes the subsystem entropy predicted by the microcanonical ensemble.  
		(b) The orange line represents the time evolution of the average mutual information between different lattice sites, defined as  $\langle I(i:j) \rangle = \frac{1}{N} \sum_{i,j} I(i:j)$  ,
		where $N = C_6^2$ is the number of site pairs. The orange shaded region indicates the standard deviation of mutual information,  $	\sigma_{I} = \sqrt{\frac{1}{N} \sum_{i,j} (I_{ij} - \langle I_{ij} \rangle)^2}$  .
		The black dashed line denotes the mutual information predicted by the microcanonical ensemble.
		The inset shows mutual information fluctuations between different lattice sites $i$ and $j$ within the time range marked by the dashed box.  
	}  
	\label{fig2}  
\end{figure}

\textit{Dynamics of entanglement entropy and mutual information}.—  
We initialize the system in an inhomogeneous particle distribution, where the first three lattice sites (1-3) are fully occupied by both types of fermions, while sites 4-6 are empty. The initial state is denoted as $
|\psi_0\rangle = |111000\rangle_\tau\otimes|111000\rangle_\upsilon$.
Throughout the evolution, the system remains in a pure state, and its density matrix is denoted as $ \rho $.  
Unlike the free propagation of a single particle in a one-dimensional lattice, the transmission of particles from sites 1-3 outward involves collisions when different types occupy the same site. These interactions lead to a rapid increase in entanglement within the system.  
To study the time evolution of the system's entanglement entropy, we partition the system into two subsystems, $ A $ and $ B $, with their reduced density matrices given by $ \rho_{A(B)} = \operatorname{Tr}_{B(A)}(\rho) $. The entanglement entropy is defined as  $S(\rho_A) = - \operatorname{Tr}(\rho_A \log \rho_A)$.
Fig.~\ref{fig2}(a) shows the dynamical evolution of the entanglement entropy for different choices of subsystem $ A $. The three curves correspond to $ A=\{1\}, \{1,2\}, \{1,2,3\} $, while the three horizontal lines represent the entropic limits predicted by the microcanonical ensemble, given by  $S(A) = -\ln\left(\frac{1}{\Omega}\right)$ , 
where $ \Omega $ is the number of microstates in the microcanonical ensemble.  
Initially, the entanglement entropy is zero, but as the system evolves, it approaches the entropic limit predicted by the microcanonical ensemble. For smaller subsystems $ A $, the entanglement entropy aligns more closely with the microcanonical prediction, implying that subsystem $ B $ acts as an effective environment for $ A $ \citep{genway_dynamics_2010}. As the size of $ A $ increases, the entanglement entropy exhibits a volume-law growth, a characteristic feature of thermalization.

Another characteristic of thermalization is the statistical independence between subsystems. To investigate this, we analyze the evolution of mutual information between different lattice sites $ i $ and $ j $. The mutual information is defined as  
$I(i:j) = S(\rho_{A=\{i\}}) + S(\rho_{A=\{j\}}) - S(\rho_{A=\{i,j\}})$.
As shown in Fig.~\ref{fig2}(b), the average mutual information $ \langle I(i:j) \rangle $ oscillates rapidly over time, and the standard deviation $ \sigma_{I} $ of $ I(i:j) $ remains relatively large. This behavior deviates from the statistical characteristics expected in thermal equilibrium, where mutual information between disjoint small subsystems should be minimal and stable.

Moreover, although the system's quantum state becomes highly spread over a certain period, the finite system size ensures that subsequently entangled particles undergo further collisions and scatterings within the limited space, leading to disentanglement and a return to the initial state.  
In general, for an isolated system of finite size with discrete energy levels, the system's dynamical evolution is periodic.

\begin{figure}[htbp]  
	\centering  
	\includegraphics[scale=1.0]{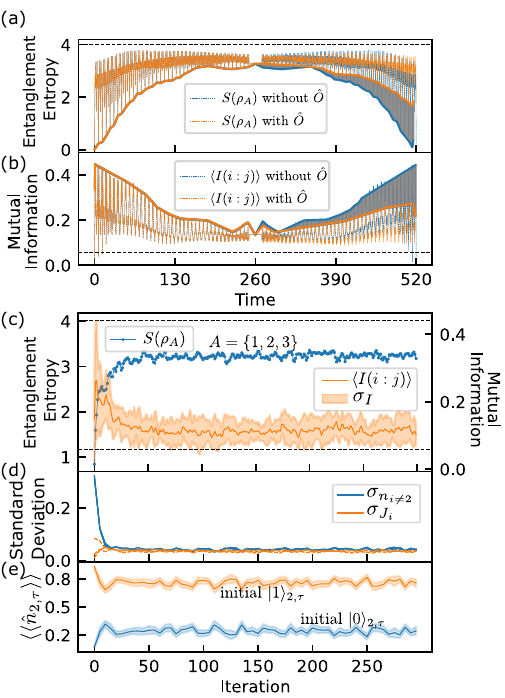}
	\caption{Simulation of Information Erasure with $\hat{O}(t)$.  
		In panels (a) and (b), the orange and blue dashed lines represent the entanglement entropy and mutual information evolution with and without local perturbations, respectively. The solid orange and blue lines are their envelopes, while the shaded areas indicate discrepancies between the two cases. The entanglement entropy $S(\rho_A)$ is computed for subsystem $A = \{1, 2, 3\}$. The system evolves forward from $t = 0$ to $250$ and backward from $t = 270$ to $520$, with local operations applied only during $t = 250$ to $270$.  
		Panel (c) depicts the evolution of entanglement entropy and mutual information under the $[\hat{O}^\dagger \hat{O}(t)]^N$ sequence, where $t = 50$, and $\hat{O}$ matches that in panels (a) and (b).  
		Panel (d) shows the standard deviation of $\langle \hat{n}_{i \neq 2} \rangle$ and $\langle \hat{J}_{i} \rangle$ over all possible Fock states as a function of $N$. Panel (e) displays the average value of $\langle \hat{n}_{2,\tau} \rangle$ for all possible Fock states at lattice site 2, with different initial states, as a function of $N$. The blue (orange) curve corresponds to the initial state $|0\rangle_{2,\tau}$ ($|1\rangle_{2,\tau}$), and the shaded region represents the corresponding statistical standard deviation.
	}
	\label{fig3}  
\end{figure}

\textit{Thermalization, irreversibility, and information erasure}.—
From the perspective of quantum dynamics, the increase in entropy during the thermalization process arises from the increase in entanglement between the system’s microscopic degrees of freedom. On the other hand, the reverse evolution process implies disentangling, which, from the perspective of wave function evolution, is a process where a highly spread-out wave function undergoes constructive interference during reverse propagation.
This process is highly susceptible to disturbance from local perturbations. To investigate the destruction mechanism of reverse evolution caused by local perturbations, we first evolve the system from the initial state for a period of time $ t_{\rightarrow} = 250 $, then insert a local operation $ \hat{O} $, and perform the reverse evolution for the same amount of time $ t_{\leftarrow} = t_{\rightarrow} $. We observe how the reverse evolution is altered when the operation $ \hat{O} $ is added. The local operation $ \hat{O} $ we used is to apply a local potential $ U_{2,\tau} = 1 $ on the $ \tau $-particle at lattice site $ i = 2 $ and evolve it for a time $ 20 $, which is equivalent to performing a local phase operation. During the action of the local operation $ \hat{O} $, the system's Hamiltonian is set to zero.
The numerical simulation results, shown in Fig.~\ref{fig3}(a), reveal the dynamical evolution of entanglement entropy and mutual information. The blue and orange solid lines represent the envelope of the dynamical evolution curves of entanglement entropy and mutual information with and without the local perturbation. It can be seen that after the insertion of the local perturbation, the direction of reverse evolution changes. At time $ t = 250 $, the initial non-equilibrium state information is preserved. The $\tau$ and $\upsilon$ particles become highly entangled and carry each other's information, while the particle at site 2 is highly entangled with other particles in the system. 
The local operation $\hat{O}$ preserves the entanglement between the particle at site 2 and the others, but induces a phase shift. Consequently, the propagation speed and direction of the particle flow at this site are modified.
This local change spreads over time, causing a greater divergence between the forward and reverse evolutions. The reverse evolution with the perturbation exhibits higher entanglement entropy and lower mutual information compared to the unperturbed evolution.

Here, we focus on the reduction of quantum state information caused by $\hat{O_i}(t_i)$. Since the isolated system remains in a pure state, its quantum state information is quantified by the entanglement entropy. A higher entropy corresponds to less information.
 In our numerical simulation, we examine the process of erasing system quantum state information during thermalization by applying a sequence $[\hat{O}^\dagger \hat{O}(t)]^N$. With fixed $\hat{O}$ and $t$ here, non-commutativity of the time evolution operators is introduced by inserting $\hat{O}^\dagger$.
As shown in Fig.~\ref{fig3}(c), applying this sequence leads to a monotonic increase in the system's entanglement entropy until it reaches a stable value. More importantly, the mutual information between different lattices in the system steadily decreases, reaching a stable value. This stable value is related to the spread time $t$ of $\hat{O}(t)$.
To illustrate the information erasure property of the $[\hat{O}^\dagger \hat{O}(t)]^N$ sequence, we examine the expectation values of the particle number and particle current at different lattice sites in the final state, considering all possible Fock states as the initial states of the system. The standard deviation of local observable expectation values across different initial states as a function of $N$ is shown in Fig.~\ref{fig3}(d). Beyond a certain $N$, the expectation values of microscopic mechanical quantities for different initial states converge, indicating that after undergoing the $[\hat{O}^\dagger\hat{O}(t)]^N$ sequence, these quantum states become indistinguishable through local measurements.
Lattice site 2, where the local operation $\hat{O}$ is applied, retains some dependence on the initial state. Specifically, if the initial state of lattice site 2 is $|0\rangle_{2,\tau}$ ($|1\rangle_{2,\tau}$), the expectation values in the final state for all possible Fock initial states converge, as shown by the blue (orange) line in Fig.~\ref{fig3}(e), with the shaded region representing the standard deviation over different initial states. It can be inferred that in a general sequence of $\hat{O_i}(t_i)$, there is no issue of initial state information retention.

\begin{figure}[htbp]  
	\centering  
	\includegraphics[scale=1.0]{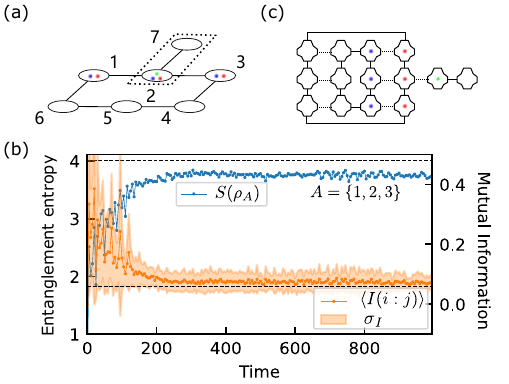} 
	\caption{Isolated system model and entanglement entropy dynamics. 
		(a) A perturbative particle $\phi$ is introduced into the model from Fig.~\ref{fig1}(b). This particle moves only between two lattice sites within the dashed box and starts at lattice site 2.  
		(b) Dynamical evolution of entanglement entropy and mutual information after introducing $\phi$. The entanglement entropy $S(\rho_A)$ is computed for subsystem $A = \{1,2,3\}$, considering only the $\tau$ and $\upsilon$ particles. In (b), the orange line represents the average mutual information between lattice sites 1–6, with the shaded area indicating its standard deviation.  
		(c) The model structure in (a) mapped onto a superconducting qubit chip. Polygons represent superconducting qubits, solid lines denote $XY$ coupling, and dashed lines indicate $ZZ$ coupling. Qubits marked with small balls are initialized in state $|1\rangle$.  
	}  
	\label{fig4}  
\end{figure}

\textit{Thermalization of an isolated many-body system}.— To incorporate the information erasure mechanism into an isolated system, we modify the lattice in the model shown in Fig.~\ref{fig1}(b) by adding a $\phi$ particle, as illustrated in Fig.~\ref{fig4}(a).
The $\phi$ particle can move only between two lattice sites within the dashed box, while the $\tau$ and $\upsilon$ particles remain restricted to the original one-dimensional lattice. When the $\phi$ particle occupies lattice site 2, it interacts with the $\tau$ particle at that site with an interaction strength $U_{\tau,\phi} = -0.7$, while $J_{\phi} = 3$. The Hamiltonians for the $\tau$ and $\upsilon$ particles remain unchanged.
In this case, the scattering collision between the newly introduced $\phi$ particle and the $\tau$ particle at lattice site 2 can be viewed as a local operation added to the original system. In the initial state, the $\tau$ and $\upsilon$ particles remain unchanged, while the $\phi$ particle is positioned at lattice site 2.
The system then evolves freely, with its entanglement entropy and mutual information dynamics shown in Fig.~\ref{fig4}(b). The entanglement entropy rapidly reaches its maximum and stabilizes. Simultaneously, as entanglement entropy increases, the mutual information between lattice sites quickly approaches the limit predicted by the microcanonical distribution. Compared to the periodic oscillations of the original system, this isolated system exhibits rapid thermalization.

\textit{Summary and outlook}.— From the perspective of quantum information, we demonstrate the mechanism of irreversibility and thermalization in the $\hat{O_i}(t_i)$ sequence through local erasure of spread non-equilibrium initial state information.
This mechanism is closely related to research on the decoherence\citep{zurek_decoherence_2003}, eigenstate thermalization hypothesis, quantum information scrambling\citep{landsman_verified_2019, yan_information_2020, touil_quantum_2020, touil_information_2024}, and information erasure\citep{landauer_irreversibility_1961, piechocinska_information_2000, groisman_quantum_2005, pati_physical_2012}, and further theoretical investigations are necessary. 
Future quantum simulations will enable unprecedented exploration of these phenomena.

\appendix
\section{Supplemental Material}
\label{supp:sec1}
\subsection{Quantum simulation on a superconducting quantum processor}
Quantum simulations enable the study of the quantum dynamical evolution of the Hubbard model. Using the model in Fig.~\ref{fig4}(a) as an example, we illustrate how such research can be conducted on a superconducting quantum chip. A 14-qubit chip can be constructed, as shown in Fig.~\ref{fig4}(c), consisting of two 6-qubit rings forming inner and outer circles, with two additional qubits connected to the outer ring.
Between neighboring qubits within each ring, there is a nearest-neighbor XY interaction, while between corresponding qubits in the inner and outer rings, there is a ZZ interaction. The other two qubits have only ZZ coupling with the qubits in the outer ring.
The $XY$ interaction term corresponds to the hopping between neighboring lattice sites in our model, while the $ZZ$ interaction represents the repulsive or attractive interaction between different particles at the same site. Interaction parameters between qubits on the superconducting qubit chip can be tuned via couplers. Additionally, the coupling strength and sign between superconducting qubits can be controlled through these couplers, enabling quantum simulation of the $e^{i\hat{H}t_i} \hat{O_i} e^{-i\hat{H}t_i}$ sequence evolution.
There have been some experimental demonstrations of dynamical evolution in one-dimensional bosonic Hubbard models on superconducting qubit chips \citep{guo_observation_2021}, as well as studies on thermalization and scrambling in bosonic Hubbard models on ladder structures \citep{zhu_observation_2022, chen_observation_2021}, and other experiments on two-dimensional bosonic Hubbard models \citep{yanay_two-dimensional_2020, karamlou_probing_2024, zhang_many-body_2023}.

%

\bibliography{irreversibility}

\end{document}